\documentstyle[prd,aps]{revtex}

\begin{document}

\title{Big-Bang Cosmology with Photon Creation}

\author{U. F. Wichoski \footnote[1]{email: 
wichoski@het.brown.edu}}
\address{Department of Physics, Brown University, \\
Providence, RI 02912, USA}

\author{J. A. S. Lima \footnote[2]{e-mail:
limajas@dfte.ufrn.br}}
\address{Departamento de F\'{\i}sica Te\'orica e Experimental, \\
Universidade Federal do Rio Grande do Norte, \\
59072 - 970, Natal - RN, Brazil}

\maketitle

\vskip 1.5cm

\begin{abstract}
\noindent 
The temperature evolution law is determined for an expanding FRW type 
Universe with a mixture of matter and radiation where ``adiabatic" 
creation of photons has taken place. Taking into account this photon 
creation we discuss the physical conditions for having a hot big bang 
Universe. We also compare our results to the ones obtained 
from the standard FRW model. 
\end{abstract}

\setcounter{page}{0}
\thispagestyle{empty}

\vfill

\noindent BROWN-HET-1089\hfill      August   1997.

\noindent astro-ph/9708215 \hfill Typeset in REV\TeX

\vfill\eject

\baselineskip 24pt plus 2pt minus 2pt

\section{Introduction}

It is widely believed that matter and radiation need to be created in 
order to overcome some conceptual problems of the standard hot big-bang 
cosmology \cite{Guth}. The most popular approach accounting for the 
phenomenon of creation is based on the idea that the early Universe 
underwent an inflationary phase during which the temperature decreased 
nearly $10^{28}$ orders of magnitude. At the end of this supercooling 
process, the energy density of the inflaton field was completely or 
almost completely converted into radiation, and the resulting Universe 
could have been reheated in less than one expansion Hubble 
time \cite{Traschen}.
However, there are theories where the gravitational particle creation 
phenomenon is conceived with no appealing for inflation and, 
consequently, allow the creation process to occur continuously in 
the course of the evolution. Probably, the best example is the 
adiabatic vacuum mechanism invented long ago by Parker and 
collaborators using the Bugoliubov mode-mixing technique in the 
context of quantum field theory in curved spacetimes 
\cite{Parker,BD}. However, this approach is plagued 
with several conceptual and mathematical difficulties. In particular, 
there is not a well-defined prescription of how the created matter 
and/or radiation should be incorporated in the Einstein field equations 
(EFE) \cite{Branden}. 

More recently, a new phenomenological macroscopic approach to 
gravitational 
creation of matter and radiation has attracted considerable attention 
\cite{Prig 89}-\cite{L 96}. In this framework, the creation event of the 
inflationary scenario is also replaced by a continuous creation process. 
The crucial ingredients of this formulation are a balance equation for 
the number density of the created particles and a negative pressure term 
in the stress tensor so that the back-reaction problem present in 
Parker's mechanism is naturally avoided. Another advantage of this 
formulation is that the laws of non-equilibrium thermodynamics were used 
since the very beginning, thereby leading to definite relations among 
the classical thermodynamic quantities. In particular, the creation 
pressure depends on the creation rate in a well defined form, and 
potentially may alter significantly several predictions of the standard 
big-bang cosmology. Completing such an approach, a spectrum for 
blackbody radiation when photon creation takes place has also been 
proposed in the literature \cite{L 96,L1 97}. This spectrum is preserved 
during a free expansion (for instance, after decoupling between matter 
and radiation), and more important still, it is compatible with the 
present spectral shape of the cosmic background radiation (CBR). 

On the other hand, in the photon-conserving 
Friedmann-Robertson-Walker (FRW) Universes, the 
temperature of the matter content follows 
the radiation temperature law 
when there is any thermal contact between 
these components. This state of affairs define what is 
called a hot big-bang Universe. Usually, the condition 
that the Universe underwent a very hot phase 
in its beginning 
is expressed by requiring that there 
are many photons for each proton or neutron in 
the Universe today. This fact allows one to 
establish the cosmic eras, and is closely related 
to the high value of the radiation 
specific entropy (per baryon) in the present Universe. 

In this letter, by taking into account the photon 
creation process described by the 
thermodynamic formulation 
of irreversible processes, 
we analyze the temperature 
evolution law for the 
matter-energy content in the framework of a FRW 
metric. Our aim here is to discuss 
under which conditions the basic concept of hot 
big bang Universe remains valid when a 
continuous photon creation phenomenon is considered. 

\section{CBR spectrum and the temperature law with photon creation}

Let us consider a spectrum of photons whose 
number and energy densities 
are, respectively, $n_r \sim T^{3}$ and 
$\rho_r \sim T^{4}$ and let $N_r(t)$ be the 
instantaneous comoving total number of 
photons, where $T$ is the temperature. 
Since $N_r=n_rR^{3}$, where 
R(t) is the scale factor of a FRW 
cosmology, one may write 
\begin{equation} 
\label{eq:AF}
N_r(t)^{-\frac {1}{3}}TR = const  \quad .
\end{equation}
It thus follows that if $N_{r}(t)$ is a constant the 
temperature law 
of the photon conserving FRW model, $TR=const$, is 
recovered. From the scale 
factor-redshift relation, $R = R_o(1 + z)^{-1}$, the 
above temperature law becomes 
\begin{equation}
\label{eq:TL}
T = T_o(1 + z)(\frac {N_r(t)}{N_{or}})^{\frac{1}{3}}
\quad ,
\end{equation}
where the subscript $o$ denotes the present day 
value of a quantity. As we shall 
see this relation has some 
interesting physical consequences, and a 
consistent cosmological framework with 
photon creation may be traced back using 
this new temperature law. 

As shown by one of us \cite{L 96,L1 97}, equation (\ref{eq:AF}) 
leads to the following spectral distribution: 

\begin{equation}
\label{eq:RO}
\rho _{T}(\nu) = (\frac {N_r(t)}{N_{or}})^\frac {4}{3}
\frac {8 \pi h}{c^{3}}
\frac {\nu ^{3}}
{exp[(\frac {N_r(t)}{N_{or}})^\frac{1}{3}
{\frac {h\nu}{kT}}]  - 1} \quad.
\end{equation}
In the absence of creation ($N_{r}(t)=N_{or}$), 
the standard Planckian spectrum is recovered\cite{SN}. 
The  derivation of the above spectrum 
depends only on the new temperature 
law and satisfies the equilibrium relations 

\begin{equation}
\label{eq:NR}
n_r (T) =\int_{0}^{\infty}
\frac {\rho_{T}(\nu) d\nu}
{(\frac {N_r(t)}
{N_{or}})^\frac {1}{3}
h\nu} =bT^{3} \quad,
\end{equation}
and

\begin{equation}
\label{RR}
\rho_r (T) =\int_{0}^
{\infty} \rho_{T}(\nu )d\nu =
aT^{4} \quad,
\end{equation}
where
$b=\frac {0.244}
{{\hbar^
{3}} c^
{3}}$
and
$a=\frac {\pi^
{2} k^{4}}
{{15\hbar^
{3}} c^{3}}$, are
the
blackbody radiation 
constants. A gravitational 
photon creation process satisfying the 
above equilibrium relations has been 
termed ``adiabatic" creation \cite{CLW 92,L 96}. 

The temperature law (\ref{eq:AF}) implies that the exponential factor 
appearing in the spectrum given by Eq.(\ref{eq:RO}) is time 
independent. As a consequence, the spectrum is not destroyed as the 
Universe evolves, at least not  after the 
transition from an opaque to a transparent Universe. Note also 
that the above distribution cannot be distinguished from the blackbody 
spectrum at the present epoch when $T=T_o$ and $N_r(t_o)=N_{or}$. 
In what follows we study under which conditions the temperature law 
(\ref{eq:AF}) may be applied before decoupling, that is, during the 
time when matter and radiation were in thermal contact. 

Let us now consider a mixture of a non-relativistic gas 
in thermal contact with the blackbody 
radiation described by Eq.(\ref{eq:RO}). For 
completeness we set up the basic equations including 
``adiabatic" creation of both components. 

For this system, the total pressure ($p$) and energy 
density ($\rho$) are given by ($c=1$) 

\begin{equation}
\label{eq:MD}
\rho = n m + (\gamma - 1)^{-1} n k T + a T^{4} =
\rho_{m} + \rho_{r} \quad ,
\end{equation}

\begin{equation}
\label{eq:MP}
p =  n k T + \frac{1}{3} a T^{4} + p_{rc} + p_{mc} \quad ,
\end{equation}
where $n$ is the number density of gas
particles, m is their mass, $\gamma$ is
the specific ratio of the
gas ($5/3$ for a monoatomic gas). The
quantities $p_{rc}$ and $p_{mc}$ stand
for radiation and matter creation
pressures, respectively.
In the ``adiabatic" formulation they
assume the following form \cite{CLW 92,LG 92}:
\begin{equation}
\label{eq:RC}
    p_{rc} = - \frac{\rho_r + p_r}{3n_{r}H} \psi_{r} \quad,
\end{equation}
and
\begin{equation}
\label{eq:MC}
    p_{mc} = - \frac{\rho_m + p_m}{3nH} \psi_m \quad,
\end{equation}
where $H = {\dot {R}}/R$ is the Hubble 
parameter and $\psi_r$, $\psi_m$ 
denote, respectively, the creation rates of 
radiation and matter components. The number 
densities, $n$ and $n_r$, satisfy 
the balance equations: 

\begin{equation}
\label{eq:BM}
      \frac{\dot{n}}{n} + 3 \frac{\dot{R}}{R} =
           \frac{\psi_m}{n}
               \quad ,
\end{equation}
and
\begin{equation}
\label{eq:BR}
      \frac{{\dot{n_r}}}{n_r} + 3 \frac{\dot{R}}{R} =
           \frac{\psi_r}{n_r}
               \quad.
\end{equation}

In the context of the FRW metric, the energy 
conservation law 
reads \cite{Weinberg} 
\begin{equation} 
\label{eq:CL}
\frac{d}{dR}(\rho R^{3}) = - 3pR^{2} \quad . 
\end{equation}

Before proceeding further, it is worth noticing that 
the analysis of the temperature evolution law may be 
separated in several 
cases: (i) the standard 
model ($\psi_r = \psi_m=0$); (ii) photon 
creation ($\psi_r \neq 0, \psi_m=0)$; (iii) matter 
creation $(\psi_r = 0, \psi_m \neq 0)$; and (iv) radiation 
and matter creation $(\psi_r \neq 0, \psi_m \neq 0)$. 
In this letter we are primarily interested in the case 
of photon creation whose spectrum is defined by Eq.(\ref{eq:RO}). 
Thus, henceforth we restrict our attention to the case (ii), 
for which $\psi_m = p_{mc} = 0$. 

Inserting Eq.(\ref{eq:MD}) and Eq.(\ref{eq:MP}) into Eq.(\ref{eq:CL}) 
it follows that 
\begin{equation} 
\label{eq:BE} 
\frac{1}{R^{2}}\frac{d}{dR} \left[ n m R^{3} + 
(\gamma - 1)^{-1} n k T R^{3} + 
a T^{4} R^{3} \right] = - 3 n k T - a T^{4} + 
\frac{4}{3} a T^{4}\frac {\psi_{r}}{n_{r}H} \quad.
\end{equation}

Now, by considering that the number of 
massive particles is conserved, one 
obtains from Eq.(\ref{eq:BM}) 
\begin{equation} 
\label{eq:PC}
\frac{d}{dR} (n R^{3}) = 0 \quad.
\end{equation}
Thus, substituting Eq.(\ref{eq:PC}) into Eq.(\ref{eq:BE}), 
and dividing the result  by $3 n k T$, we get 
\begin{equation} 
\label{eq:TR}
\frac{R}{T} \frac{dT}{dR} = -\frac {1 + \sigma_{r} -\sigma_{r}{\beta}}
{\frac{1}{3}(\gamma - 1)^{-1} + \sigma_r} \quad ,
\end{equation}
where the parameters $\sigma_r$ and $\beta$ are defined by 
\begin{equation}
\label{eq:SI}
\sigma_r= \frac{4aT^{3}}{3nk} \quad,
\end{equation}
and 
\begin{equation}
\label{eq:BT}
\beta = \frac{\psi_{r}}{3n_{r} H} \quad.
\end{equation}
The quantity $\sigma_r$ 
is the specific radiation entropy 
(per gas particle) while $\beta$ 
quantifies how relevant is the photon 
creation process in an 
expanding Universe. 
These parameters are dimensionless, and in 
general, are also time-dependent 
quantities. For $\beta=0$, equation (\ref{eq:TR}) 
reduces to that one of the 
standard model \cite{Weinberg}. It is worth noticing that the 
$\beta$ parameter may be rewritten as $\beta=\frac{\Gamma}{H}$, 
where $\Gamma=\frac{\psi_r}{3n_r}$. Thus, $\beta$ 
is the ratio between the ``interaction rate" of the creation 
process and the 
expansion rate of the Universe. Naturally, $\Gamma$ or 
equivalently $\psi_r$, must be determined from a kinetic 
theoretical approach or from a 
quantum field theory. In any case, a reasonable upper limit 
to this rate is $\Gamma=H$, since for this value the photon 
creation rate exactly compensates for the 
dilution of particles due to expansion (see Eq.(\ref{eq:BR})). 
A new cosmological scenario will be thus 
obtained only if $0<\beta\leq1$. In particular, for a 
radiation dominated model ($p=\frac{1}{3}\rho$), 
the limiting case $\beta=1$ is a 
radiation-filled de Sitter Universe. Naturally, if 
$\beta$ is small, say, if 
$\Gamma \simeq 10^{-3}H$ or smaller, 
for all practical purposes the photon creation 
process may be safely neglected. 

Returning to equation (\ref{eq:TR}) we see that in the limit 
$\sigma_{r} \gg 1$, the 
temperature evolution equation reduces to 
\begin{equation}
\label{eq:RTB}
\frac{R}{T} \frac{dT}{dR} = - (1 - \beta) \quad ,
\end{equation}
or still, using Eq.(\ref{eq:BR}) and Eq.(\ref{eq:BT}) 
\begin{equation}
\label{eq:RTN}
\frac{R}{T} \frac{dT}{dR}=
\frac{\dot n_r}{3n_rH}\quad.
\end{equation}
This equation can be rewritten as 
\begin{equation}
\label{eq:TRN}
\frac{dT}{T}= \frac{dN_r}{3N_r} - \frac{dR}{R} \quad,
\end{equation}
where $N_r(t)$ is the instantaneous comoving number of photons. 
Integrating the above expression we obtain: 
\begin{equation}
\label{eq:AF1}
N_r(t)^{-\frac {1}{3}}TR = const  \quad ,
\end{equation}
which is the same temperature law 
for a freely propagating blackbody spectrum with 
photon creation (see Eq.(\ref{eq:AF})). 

Therefore, as long as the quantity $\sigma_r$ is 
large, the radiative component will continue to 
overpower the material component. Hence, while 
there is any significant thermal 
contact between them, the matter temperature will 
follow Eq.(\ref{eq:AF1}) as well. This means that 
the condition for a hot big bang cosmology 
is not modified 
when ``adiabatic" photon creation 
occurs. Note that 
the above result holds regardless of the value 
of the $\beta$ parameter and 
assures the validity of the 
above equation during a 
considerable part of the 
evolution of the 
Universe. The important point here is that the 
cosmic eras are still viable using the above 
generalized temperature law.  However, unlike 
in the standard model, the radiation specific 
entropy does not remain 
constant when photon creation takes place. Since 
$n_r \propto T^{3}$ the usual expression 
\begin{equation}
\label{eq:SRN}
\sigma_r=0.37\frac{n_r}{n}=
0.37\frac{N_r(t)}{N}
\end{equation}
remains valid nonetheless $n_r$ does not vary proportionally 
to $R^{-3}$. As a consequence, the variation 
rate of $\sigma_r$ is directly proportional to 
the variation rate of $N_r(t)$, which in turn 
depends on the magnitude of the $\beta$ parameter. 
So, if $\beta$ approaches to zero, $N_r(t)$ and 
$\sigma_r$ assume their constant values, and the 
standard model results are recovered. 

\section{A specific model}

We consider the simplest  
photon creation model for which the 
parameter $\beta$ is  constant. This scenario can be defined by taking 
into account 
the ``interaction" rate $\Gamma= \alpha H$, where $\alpha$ is a positive 
constant smaller than unity. As one may check from Eq.(\ref{eq:RTB}), 
in this case the temperature 
scaling law assumes the simple form 
\begin{equation} 
\label{eq:BCT}
T \propto R^{- (1 - \alpha)}\quad,
\end{equation}
or still, in terms of the redshift
\begin{equation}
\label{eq:BCTZ}
T = T_o (1 + z)^{1 - \alpha}\quad,
\end{equation}
so that for $z>0$ the Universe is cooler than the standard model. 

Considering the balance equation written as 
\[
\dot{n_{r}} + 3 (1 - {\alpha}) n_r H = 0 \quad ,
\]
we obtain, 
\begin{equation}
\label{eq:NRB}
n_{r} \propto R^{-3(1-\alpha)} \quad ,
\end{equation}
which could be obtained directly from the 
number density-temperature relation 
(see Eq.(\ref{eq:NR})). As a consistency check, by substituting 
$N_r=n_rR^{3}$ in the expression above, it is easily seen 
that the temperature law also follows directly from the generalized 
expression Eq.(\ref{eq:AF1}). 
Replacing Eq.(\ref{eq:NRB}) 
into Eq.(\ref{eq:SRN}) one may see that 
\begin{equation}
\label{eq:SRB}
\sigma_r= {\sigma_{or}}(\frac{R}{R_o})^{3\alpha} \quad,
\end{equation}
or still, in terms of the redshift 
\begin{equation}
\label{eq:SRD}
\sigma_r=\frac {\sigma_{or}}{(1 + z)^{3\alpha}} \quad,
\end{equation}
where $\sigma_{or} \sim 10^{8-9}$ is the now observed 
radiation specific entropy. 

The above equations are a concrete 
example that the usual 
physical conditions defining a hot 
big bang cosmology may be weakened. In particular, 
specific entropy so large (and constant) as $10^{8}$ 
is not 
required, providing that the product $TR$ varies 
in the course of the evolution. In other 
words, $10^{8}$  is only the 
present value of an increasing 
time-dependent quantity. Therefore, instead of predicting 
the currently observed value, the 
important question in this framework is how a 
reasonable ``initial value'', say 
$\sigma_r \simeq 10^{2}$, could be 
explained from the first principles. In 
particular, for the toy model presented here, it 
follows from equations (\ref{eq:BCTZ}) and (\ref{eq:SRD}), 
that a value of $\sigma_r \simeq 10^{2}$ in the beginning of 
the nucleosynthesis epoch is 
possible only if $\alpha \simeq 0.18$. More details on the 
nucleosynthesis of light elements, using a properly 
modified nucleosynthesis code which considers 
``adiabatic" creation of neutrinos and effectively massless 
species at nucleosynthesis epoch, will be discussed in a 
forthcoming communication \cite{AJU}. 

In conclusion, we have considered an 
evolutionary Universe where ``adiabatic" photon 
creation has taken place. The conditions defining a 
big bang scenario with a blackbody spectrum endowed with 
photon creation and 
compatible with the present observed CBR distribution 
have been discussed. As we know, earlier approaches to 
matter creation processes, for instance, the steady state 
model, C-field theory, scale-covariant theory 
and others\cite{GS 78}, fail the test of the CBR spectrum. 
As we have seen, this does not happen with the thermodynamic 
approach considered here. 
Naturally, in order to have a viable 
alternative to the photon-conserving FRW model, other 
cosmological properties need to be investigated. 
In particular, it would be important to use the 
Sachs-Wolf effect to test a big bang model with ``adiabatic" 
photon creation. 

\section*{Acknowledgments}

It is a pleasure to thank Robert Brandenberger and Jackson Maia for 
their valuable comments. This work was partially supported by 
the US Department of Energy under grant DE-F602-91ER40688, Task A, 
(at Brown), and by Conselho Nacional de Desenvolvimento Cient\'{\i}fico e 
Tecnol\'{o}gico - CNPq (Brazilian Research Agency), (JASL).


\begin{references}
\bibitem{SN} The ``adiabatic'' creation of neutrinos may also be easily 
described by  the same formalism. By choosing units such that 
$\hbar=k=c=1$, the spectral distribution for a massless gas with $g$ 
internal degrees of freedom can be written as $\rho_{T}(\omega) = (\frac
{N_r(t)}{N_{or}})^{\frac {4}{3}}\frac
{g}{2{\pi}^{2}}\omega^{3}\left[(\frac
{N_r(t)}{N_{or}})^{\frac{1}{3}}{\frac {\omega}{T}} - 
\varepsilon\right]^{-1}$, where the number $\varepsilon$ is $+1$ for 
photons and $-1$ for fermions. As expected, at early times, the 
thermal radiation 
energy density due to the relativistic particles at temperature 
$T$ is given by the usual expression $\rho = g_{*}\frac
{\pi^{2}}{30}T^{4}$, where $g^{*}(T)$ counts the total number of 
effectively massless degrees of freedom of all the relativistic particle 
species \cite{KT}.
\end{references}
\end{document}